\documentclass[prl,twocolumn,superscriptaddress,floatfix,showpacs]{revtex4}
\usepackage{amsmath,amsfonts,epsf,dcolumn,natbib,graphicx}
\usepackage{ascii}
\usepackage{xcolor}
\usepackage{soul}
\usepackage{ulem}

\newcommand{\be}{\begin{equation}}
\newcommand{\ee}{\end{equation}}
\newcommand{\beq}{\begin{equation}}
\newcommand{\eeq}{\end{equation}}
\newcommand{\bea}{\begin{eqnarray}}
\newcommand{\eea}{\end{eqnarray}}

\begin{document}
\bibliographystyle{plainnat}
%
\title{
Comment on \lq\lq Communication: Simple and accurate uniform electron gas 
correlation energy 
for the full range of densities" 
[J.\ Chem.\ Phys. \textbf{145}, 021101 (2016)]
}

\author{Valentin V.~Karasiev}
\email{vkarasev@qtp.ufl.edu}
\affiliation{Quantum Theory Project, 
Departments of Physics and of Chemistry,
University of Florida, Gainesville FL 32611-8435, U.S.A.}

\date{October 06, 2016}
\begin{abstract}
\noindent

\end{abstract}


\maketitle

Recently Chachiyo \cite{Chachiyo.2016} presented an elegant and 
simple expression for 
the uniform electron gas (UEG) correlation energy applicable over 
the full range of densities. The functional has the form 
\be
\varepsilon_{\mathrm{c}}(r_{\mathrm{s}})=a\ln(1+\frac{b_1}{r_{\mathrm{s}}}+\frac{b}{r_{\mathrm{s}}^2})
\,,
\label{Ec}
\ee
with $b_1=b$ and $r_{\mathrm{s}}$ the usual Wigner-Seitz radius. 
Two sets of two parameters $\{a$, $b\}$ in Eq.\ (\ref{Ec})
were defined for the spin-unpolarized and fully polarized cases
from the requirement that Eq. (\ref{Ec}) approach 
the correlation energy of the UEG at high density (low $r_{\mathrm{s}}$) 
\cite{Loos.Gill.2011}.
The Chachiyo functional provides a root-mean-squared error 
with respect to the quantum Monte-Carlo (QMC) data \cite{Ceperley.Alder.1980}
which is a bit smaller than that provided by the Vosko-Wilk-Nusair 
fit \cite{VWN.1980}.

\begin{figure}
  \includegraphics*[width=7.5cm]{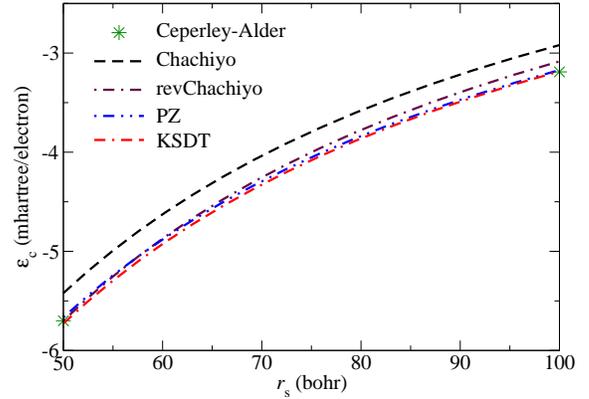}
 \caption{
Comparison among different functionals for correlation energy at large-$r_{\mathrm{s}}$ 
for spin-unpolarized UEG and reference QMC data.
}
\label{Ec-rs}
\end{figure}

Nevertheless, closer inspection reveals that the function deviates 
from the QMC data at large $r_{\mathrm{s}}$ (see Fig.\ \ref{Ec-rs}).  
The relative error is equal to 8.5\% at $r_{\mathrm{s}}=100$ bohr for 
the spin-unpolarized case. The  
corresponding relative error in the {\it total energy} per electron 
is 3.5\%
(the reference total energy components are $t_{\mathrm{s}}^{\mathrm{TF}}=0.11$, $\varepsilon_{\mathrm{x}}^{\mathrm{Slater}}=-4.58$,
$\varepsilon_{\mathrm{c}}^{\mathrm{Ceperley-Alder}}=-3.19$, all in mhartree).

The $a$ and $b$  parameters define the small-$r_{\mathrm{s}}$ 
behavior, while the term with $b_1$ contributes in the 
large-$r_{\mathrm{s}}$ regime.
This means that if we define the $b_1$ parameter from a requirement 
to match the large-$r_{\mathrm{s}}$
QMC data, the problem of discrepancies at large-$r_{\mathrm{s}}$ may be 
resolved.  The intermediate- and small-$r_{\mathrm{s}}$ behavior should 
remain essentially unchanged for small changes of $b_1$. Hence we propose 
to determine the $b_1$ parameter from the requirement
that the correlation energy Eq. (\ref{Ec}) for $r_{\mathrm{s}}=50$ bohr is equal to the QMC value. 
That gives  the new value, $b_1=1.062717673 \times b=21.7392245$ for the 
spin-unpolarized case 
and $b_1=1.034121079 \times b=28.3559732$ for the fully-polarized case.

Figure \ref{Ec-rs} shows that Eq. (\ref{Ec}) with the new $b_1$ parameter
values provides much better agreement
with the QMC data at large-$r_{\mathrm{s}}$.  There is no 
deterioration of the functional 
quality at intermediate- and small-$r_{\mathrm{s}}$. 
Mean absolute relative error (MARE), maximum relative error (MAX)
and mean absolute error (MAE) values for the original 
and revised set of parameters are shown in Table \ref{tab:table1}. For comparison the Table also includes 
data for the popular Perdew-Zunger (PZ) \cite{PZ81} and the more 
recently parametrized
Karasiev-Sjostrom-Dufty-Trickey (KSDT) \cite{LSDA-PIMC} {\it free-energy} 
functional evaluated at zero-temperature. 
Reference \cite{Sun.Perdew.Seidl.PRB.81.2010} shows that
interpolation between known high- and low-density analytic limits
can be used for accurate predictions of the correlation energy
over the whole range of densities ($0\le r_{\mathrm{s}} < \infty$),
without QMC input.
Of necessity they are significantly more
complicated than Eq.\ (\ref{Ec}).   
PZ parameters were fitted to the Ceperley-Alder QMC data \cite{Ceperley.Alder.1980}.
This functional behaves almost identically with the VWN parametrization but is
more widely used, especially in solid-state codes.
A subset of parameters in the KSDT functional was fitted to recent QMC results \cite{SND.2013}.  
The revised parameter set should make Chachiyo's simple 
expression even more generally useful than the original.  
{\it Acknowledgments:} This work was supported by U.S.\ Dept.\ of Energy grant DE-SC0002139.
Thanks to Jim Dufty and Sam Trickey for helpful discussions.
 
\begin{table}
\caption{\label{tab:table1}
MARE (\%), MAX (\%) and MAE (hartrees) for the spin-unpolarized (u) and fully-polarized (p)
UEG calculated with different functionals. All errors
are calculated w/r to the Ceperley-Alder QMC data 
in the range $2 \le r_{\mathrm{s}}\le 100$.
}
\begin{ruledtabular}
\begin{tabular}{lcccc}
\footnotesize Funct. &  \footnotesize MARE/MAX(u) & \footnotesize MAE(u) &  \footnotesize MARE/MAX(p) & \footnotesize MAE(p) \\
\hline
\footnotesize Ref.\ \cite{Chachiyo.2016}    &  3.7/8.5 & 5.4$\cdot 10^{-4}$ & 2.9/8.3 & 1.7$\cdot 10^{-4}$ \\
\footnotesize Eq. (\ref{Ec})                &  1.8/3.3 & 3.2$\cdot 10^{-4}$ & 2.8/5.6 & 2.1$\cdot 10^{-4}$ \\
\footnotesize PZ                            &  0.4/0.7 & 5.2$\cdot 10^{-5}$ & 0.5/1.0 & 3.7$\cdot 10^{-5}$ \\
\footnotesize KSDT                          &  0.9/1.5 & 1.9$\cdot 10^{-4}$ & 2.0/5.7 & 1.4$\cdot 10^{-4}$ \\
\end{tabular}
\end{ruledtabular}
\end{table}


\end{document}